\documentclass[10pt,showpacs,preprintnumbers,amsmath,amssymb]{revtex4}

\usepackage{graphicx}
\usepackage{bm}
\usepackage{color}
\def\slashchar#1{\setbox0=\hbox{$#1$}
   \dimen0=\wd0 \setbox1=\hbox{/} \dimen1=\wd1
   \ifdim\dimen0>\dimen1 \rlap{\hbox to \dimen0{\hfil/\hfil}} #1
   \else  \rlap{\hbox to \dimen1{\hfil$#1$\hfil}} / \fi}

\begin{document}
\title{Weak Quasielastic Production of Hyperons}
\author{M. Sajjad Athar, F. Akbar, M. Rafi Alam, S. Chauhan and S. K. Singh}
\affiliation{Department of Physics, Aligarh Muslim University, Aligarh-202 002, India}

\begin{abstract}
 We present the results for 
 antineutrino induced quasielastic hyperon production from nucleon and nuclear targets~\cite{Alam:2014bya,Singh:2006xp}. 
 The inputs are the nucleon-hyperon(N--Y) transition form
factors determined from the analysis of neutrino-nucleon scattering and semileptonic decays
of neutron and hyperons using SU(3) symmetry. The
calculations for the nuclear targets are done in local density approximation. The nuclear medium
effects(NME) like Fermi motion, Pauli blocking and final state interaction(FSI) effects due to 
hyperon-nucleon scattering have been taken into account. 
The hyperons giving rise to pions through weak decays also contribute to the weak pion production in addition to the $\Delta$ excitation mechanism which dominates 
in the energy region of $<$ 0.7 GeV. We also present the results of longitudinal and perpendicular components of polarization of final 
hyperon~\cite{Akbar:2016awk}. These measurements in the future accelerator experiments with antineutrinos may give some information 
on axial vector and  pseudoscalar form factors in the strangeness sector.
\end{abstract}

\maketitle
\section{Introduction}
In the few GeV energy region, experiments using neutrino as well as antineutrino beams on nuclear targets are being performed to measure precisely some of the 
 parameters of PMNS matrix, determination of CP violation in the leptonic sector, etc. (Anti)neutrino cross section in this energy region is difficult to simulate as the contribution to the cross 
 section comes from various scattering processes like quasielastic, one pion and multi pion production, 
 associated particle production, jet of hadrons production in the deep inelastic scattering region, etc. For quasielastic(QE) process induced by 
 antineutrinos there is an additional QE channel where a lepton is produced from nucleon target through the reaction 
 $\bar\nu_l + N \rightarrow l^+ + Y$($Y=\Lambda,~\Sigma$) i.e. a hyperon is produced in the final state. This channel
 is not allowed in the neutrino sector due to $\Delta S~=~ \Delta Q$ selection rule. This process although Cabibbo suppressed has lower energy threshold
 (0.19 GeV for $\bar\nu_e p \to e^+ \Lambda$ and 0.32 GeV for $\bar\nu_\mu p \to \mu^+ \Lambda$), in comparison to strangeness conserving associated 
 particle production process
 (0.9 GeV for $\bar\nu_e p \to e^+ YK$ and 1.1 GeV for $\bar\nu_\mu p \to \mu^+ YK$).
 In the case of inelastic channels, 
 the first reaction that 
 is considered is 1$\pi$ production(0.15 GeV for $\bar\nu_e p \to e^+ N\pi$ and 0.28 GeV for $\bar\nu_\mu p \to \mu^+ N\pi$), the cross section for which is studied by many authors. 
 This channel very well dominates over all the other possible inelastic channels when free nucleon target is considered. However, when this channel is considered 
 for bound nucleons, it has been found that due to renormalization of $\Delta$ properties in the nuclear medium and the final state interaction(FSI) of pions with 
 the residual nucleus the cross section reduces considerably, and over all the reduction could be  around 40-50$\%$ in the energy region of $E_{\bar\nu_l}~<~$1.2GeV
 ~\cite{Athar:2007wd}.
 The hyperons produced in QE reaction also give rise to pions. The branching ratios are for $\Lambda \rightarrow p\pi^-(63.9 \%),~n\pi^0(35.8 \%);~
 \Sigma^- \rightarrow n\pi^-(99.85 \%);~\Sigma^0 \rightarrow \Lambda\gamma(100 \%)$. In the case of hyperon production, 
 nuclear medium effects(NMEs) like Fermi motion and Pauli blocking have little effect on modifying the cross section from free nucleon case. However, the final state interaction of hyperons in the nucleus 
 through charge exchange hyperon nucleon scattering results in an enhancement in the $\Lambda$ production($\sim ~10\%$ in $^{40}Ar$), and 
 reduction in the $\Sigma$ production($\sim ~15\%$ in $^{40}Ar$). The decay width of pionic modes 
 of hyperons is highly suppressed in the 
 nuclear medium, making the hyperons live long enough to 
 pass through the nucleus before decaying into pions. Therefore, these pions are less affected by FSI of pions within the nuclear medium.
 
 Experimentally, there is a possibility to study the polarization of hyperons at present
facilities at Fermilab~\cite{fermi}  and J-PARC~\cite{JPARC} where high intensity beams of antineutrino are available. 
 The experiments planned with liquid argon TPC (LArTPC) detectors at
MicroBooNE~\cite{Chen:2007ae}  and the proposed SBND~\cite{Szelc:2016rjm} and DUNE experiments~\cite{Acciarri:2015uup} at
Fermilab will be able to see charged hadrons in coincidence, thus making it possible to measure polarization of hyperons by measuring the asymmetry in the angular
distribution of decaying pions, in addition to the 
cross section measurements being done at MINER$\nu$A~\cite{Wilkinson:2016wmz}. It is, therefore, most 
appropriate time to theoretically perform the calculations for the polarization observables in the Standard Model using Cabibbo  theory and/or quark models, 
using the present state of knowledge
 about the symmetry of weak hadronic currents and the properties of transition form factors associated with the matrix element between the hadronic states.
 
 The study of single hyperon production($|\Delta S|$ = 1) is interesting because
 \vspace{-3mm}
 \begin{itemize}
  \item they enable us to test the
SU(3) symmetry in our understanding of strangeness changing weak
processes.

\vspace{-3mm}

\item they provide an opportunity to measure
N-Y transition form factors which are presently known only at low $Q^2$ from hyperon semileptonic decays(HSD).

\vspace{-3mm}

\item precise predictions of $\bar\nu$-A cross section in 0.3 GeV - 1 GeV energy region are possible.

\vspace{-3mm}

\item polarization observables, like longitudinal ($P_L(Q^2)$) and perpendicular ($P_P(Q^2)$) components of polarization may give information independent of the 
cross section measurements on the axial dipole mass 
$M_A$, electric neutron form factor $G_E^n(Q^2)$ and may be sensitive to pseudoscalar form factors. 

\vspace{-3mm}

\item polarization observable $P_T(Q^2)$ in a direction perpendicular to the reaction plane can be used to make studies of T--violation in high energy weak interactions.
 \end{itemize}
 In this paper, we have discussed antineutrino induced  quasielastic charged current hyperon production on free and bound nucleons. 
 The N--Y transition form factors are determined from the
experimental data on quasielastic $(\Delta S =0)$ charged current (anti)neutrino--nucleon 
scattering and the semileptonic decay of neutron and hyperons  assuming 
G--invariance, T--invariance and SU(3) symmetry. The vector transition form factors are obtained in terms of nucleon 
electromagnetic form factors.
 A dipole parameterization for the axial vector form factor and the pseudoscalar transition form factor 
derived in terms of axial vector form factor assuming PCAC and GT relation 
extended to strangeness sector have been used in numerical evaluations. The nuclear medium
effects(NME) due to Fermi motion and final state interaction(FSI) effect due to 
hyperon-nucleon scattering have been taken into account. Furthermore, we have also presented the results for 
 longitudinal and perpendicular  components of polarization of final 
hyperon($\Lambda$,$\Sigma$) produced in the antineutrino induced quasielastic charged 
current reactions on nucleon targets.
 The details are given in
 Ref.\cite{Alam:2014bya,Singh:2006xp,Akbar:2016awk}.
\vspace{-5mm}
\section{Quasielastic Production of Hyperons}
\vspace{-3mm}
\subsection{Formalism}
We consider the following processes
\begin{eqnarray}\label{hyp-rec}
{\bar\nu_\mu}(k) + N(p)\rightarrow \mu^+(k^\prime) + Y(p^\prime);~N=p,n;~Y=\Lambda,~\Sigma^0, \Sigma^-
\end{eqnarray} 
where $k(k^\prime)$ and $p(p^\prime)$  are the momenta of initial(final) lepton and nucleon. 
The differential scattering cross section is given by,
\begin{equation}
\label{crosv.eq}
d\sigma=\frac{1}{(2\pi)^2}\frac{1}{4E_{\bar \nu} M}\delta^4(k+p-k^\prime-p^\prime)
\frac{d^3k^\prime}{2E_{k^\prime}}\frac{d^3p^\prime}{2E_{p^\prime}}\sum \overline{\sum} |{\cal{M}}|^2 ,
\end{equation}
with
\begin{equation}\label{lephad}
\quad {\cal{M}} = \frac{G_F}{\sqrt{2}}\sin\theta_c \left[\bar v(k^\prime) \gamma^\mu (1+\gamma_5) v(k)\right]~\left[\langle Y(p^{\prime})|V_{\mu} - A_{\mu}|N(p) \rangle\right].
\end{equation}
where
\begin{eqnarray}\label{vec}
 \langle Y(p^{\prime})|V_{\mu}|N(p) \rangle &=& {\bar{u}_Y}(p^\prime)\left[\gamma_\mu f_1^{NY}(q^2)+i\sigma_{\mu\nu} 
\frac{q^\nu}{M+M_Y} f_2^{NY}(q^2) +
\frac{f_3^{NY}(q^2)}{M + M_Y} q_\mu \right]u_N(p), \\
 \langle Y(p^{\prime})|A_{\mu}|N(p) \rangle &=& {\bar{u}_Y}(p^\prime)\left[\gamma_\mu \gamma_5 g_1^{NY}(q^2) + 
 i \sigma_{\mu\nu}\gamma_5 \frac{q^\nu}{M+M_Y} g_2^{NY}(q^2) +
 \frac{g_3^{NY}(q^2)} {M + M_Y} q_\mu \gamma_5 \right]u_N(p).~~~\quad
\end{eqnarray}
The form factors $f_i^{NY}(q^2)$ and $g_i^{NY}(q^2)$ are determined using  T--invariance, G--invariance, SU(3) symmetry i.e.
symmetry properties of weak currents, CVC and PCAC hypothesis. 
These symmetry considerations yield~\cite{Singh:2006xp}:
 $f_i^{NY}(q^2)$ and $g_i^{NY}(q^2)$ to be real and some relations between them i.e. ; $f_3^{NY}(q^2) = g_2^{NY}(q^2) = 0 $; ~$g_3^{NY}(q^2) = \frac{m_\pi}{m_\pi^2 - q^2} g_1^{NY}(q^2) $;~
 $f_{i=1,2}^{p\Sigma^0} \left( g_{i=1,2}^{p\Sigma^0}\right)  = 
  \frac{1}{\sqrt2} \;\; f_{i=1,2}^{n\Sigma^-} \left( g_{i=1,2}^{n\Sigma^-}\right)  $.
  
  The form factors, $f_{1,2}^{n\Sigma^-} (q^2) = (f_{1,2}^p(q^2) + 2 f_{1,2}^n(q^2)) $ and 
    $f_{1,2}^{p\Lambda} (q^2) = -\sqrt{\frac32} f_{1,2}^p(q^2)$ respectively for transitions $n \rightarrow \Sigma^- $ and
 $p \rightarrow \Lambda $. Similarly, $g_1^{n\Sigma^-}(q^2)$ is $\frac{D-F}{D+F} g_A^{np}(q^2)$ for 
 $n \rightarrow \Sigma^- $ and $-\frac{D+3 F}{\sqrt{6} (D+F)} g_A^{np}(q^2)$  for $p \rightarrow \Lambda $ transitions. We have used $D=0.804$, $F=0.463$ and 
  $g_A^{np}(q^2) = g_{A}^{np}(0) \left(1 - \frac{q^2}{M_{A}^{2}}\right)^{-2} $ 
with $g_{A}^{np}(0) =1.267$ and $M_{A} = 1.02$GeV.  
 \begin{figure}\centering
\includegraphics[height=0.22\textheight,width=0.4\textwidth]{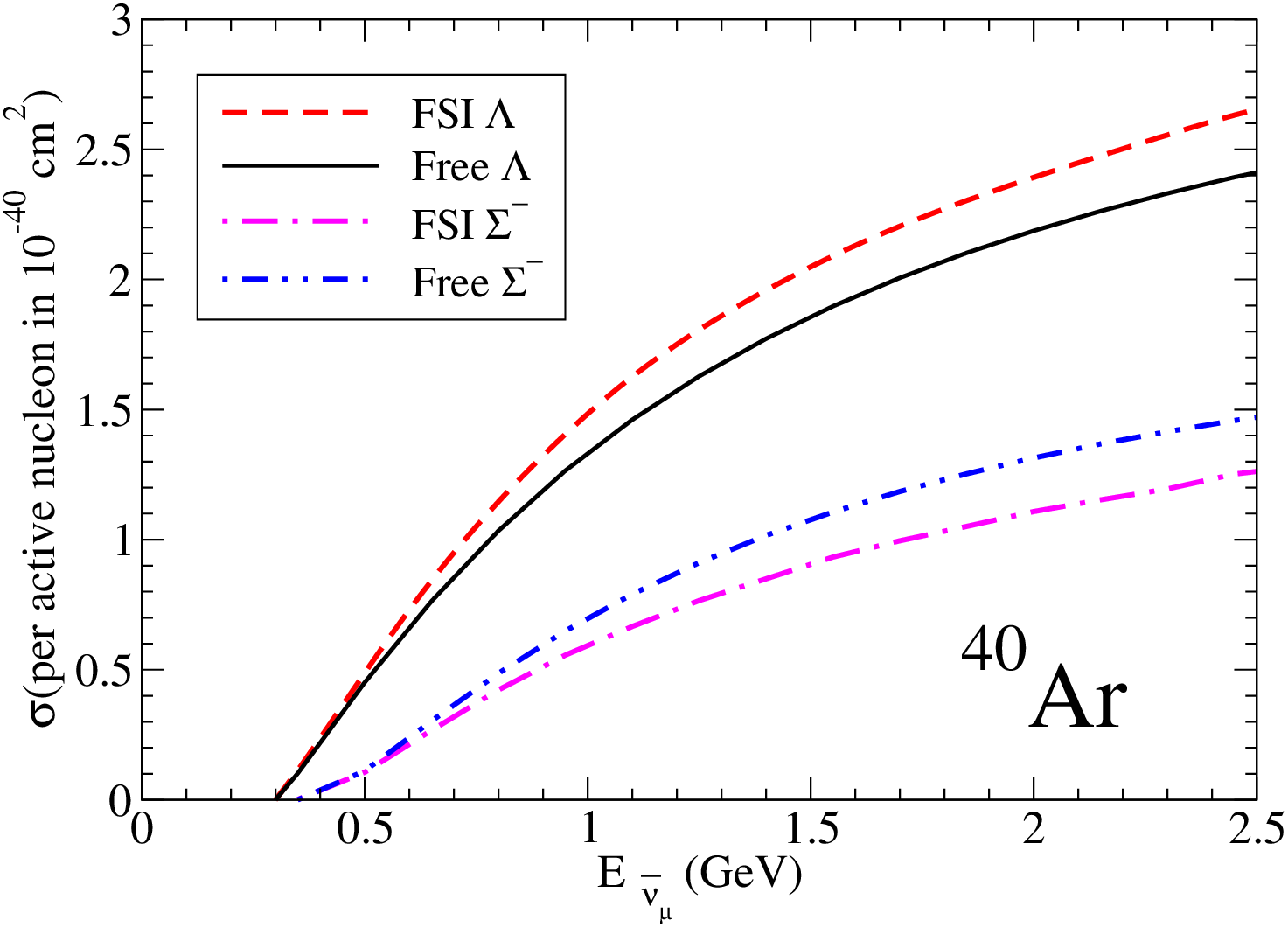}
\includegraphics[height=0.22\textheight,width=0.4\textwidth]{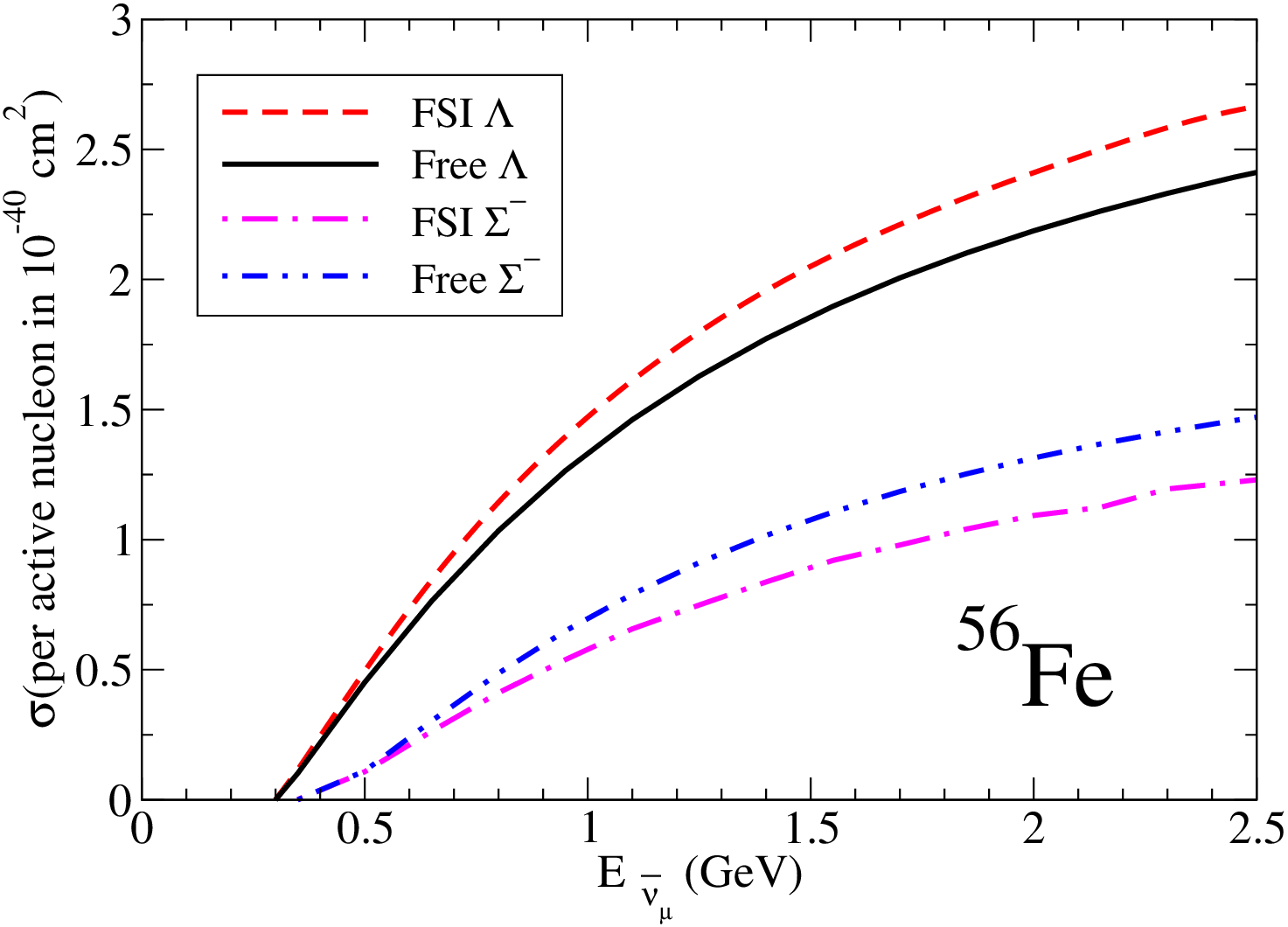}

\caption{(color online). $\sigma$(per active nucleon) vs $E_{\bar\nu_\mu}$ in $^{40}$Ar(LHS) and $^{56}$Fe(RHS), for QE hyperon production.}
\label{fg:xsec_ar}
\end{figure}

\begin{figure*}
\includegraphics[height=0.22\textheight,width=0.9\textwidth]{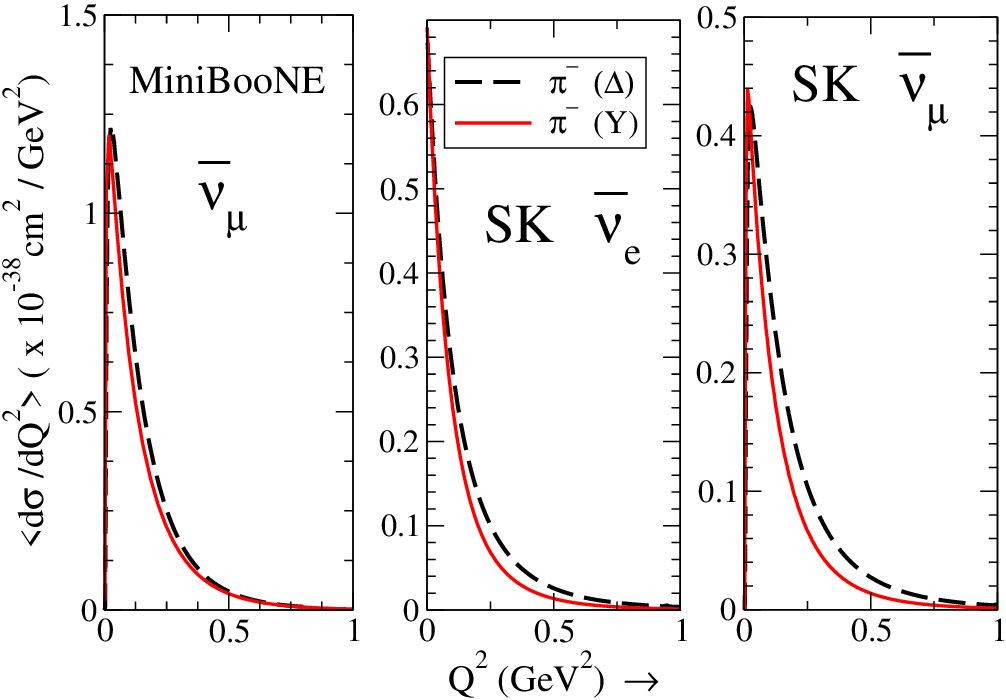}
\caption{$Q^2$ distribution for $\bar\nu_\mu$ induced reaction in $^{12}$C averaged over the MiniBooNE flux [12]
and for $\bar\nu_{e,\mu}$ induced reaction in $^{16}$O averaged over the Super-Kamiokande flux [13] are shown with nuclear medium and FSI effects. 
The $\pi^-$ production from hyperon excitations has been scaled by a
factor of 2.5.}
\label{difel.fig}
\end{figure*}

\subsection{Nuclear Medium and Final State Interaction Effects}

When these reactions take place on bound nucleons in nuclear medium, Fermi motion and 
Pauli blocking effects of nucleons are to be considered. 
In the final state after hyperons are produced, they may undergo strong interaction 
scattering processes through charge exchange( $\Sigma^- p \rightarrow \Lambda  n $, 
$\Lambda p \rightarrow \Sigma^+ n$, etc.) and inelastic ($\Lambda N \rightarrow \Sigma^0 N $,
$Y N \rightarrow Y^\prime N^\prime $)
reactions and changing the relative yield of $\Sigma^0,\; \Sigma^- $ and $\Lambda$ 
produced in the initial reactions shown in Eq.~\ref{hyp-rec}. 
In a special case, $\Sigma^+$ will appear as a result of final state interaction which are initially not produced 
through $\bar \nu_\mu N \rightarrow \mu^+ N$ reaction due to $\Delta Q = \Delta S$ rule.

The nuclear medium effects are calculated in a relativistic Fermi Gas model 
using local density approximation and the nuclear cross section is written as 
\begin{equation}\label{diffnuc}
\frac{d\sigma}{d\Omega_ldE_l}=2{\int d^3r \int \frac{d^3p}{{(2\pi)}^3}n_N(p,r)
\left[\frac{d\sigma}{d\Omega_ldE_l}\right]_{free}},
\end{equation}
where $n_N(p,r)$ is local occupation number of the initial 
nucleon of momentum $p$ and is 1 for $p < p_{F_N}$ and 0 otherwise with 
${p_F}_n={[3\pi^2\rho_n(r)]}^{1/3} \qquad \qquad \rm{for} \qquad N=n,p$.
The final state interaction of hyperon-nucleon system is calculated 
in a Monte Carlo simulation approach. 
In this approach an initial hyperon produced at a position ${\bf r}$ within the nucleus which interacts with 
a nucleon to produce a new hyperon-nucleon state $f=Y_f N_f$, within a distance $l$ with 
probability $P_Y \, dl$ where $P_Y$ is the probability per unit length given by
\begin{equation}
 P_Y=\sigma_{Y+n \rightarrow f}(E)~\rho_{n}(r)~+~\sigma_{Y+p \rightarrow
f}(E)~\rho_{p}(r),
\end{equation}
where $\rho_{n}(r)(\rho_{p}(r))$ is the local density of neutron(proton) in the
nucleus and $\sigma$ is the  cross section for $YN\rightarrow f$ process. 
 The details are given in Ref.~\cite{Singh:2006xp}.
\begin{figure}
   \includegraphics[height=5cm,width=5cm]{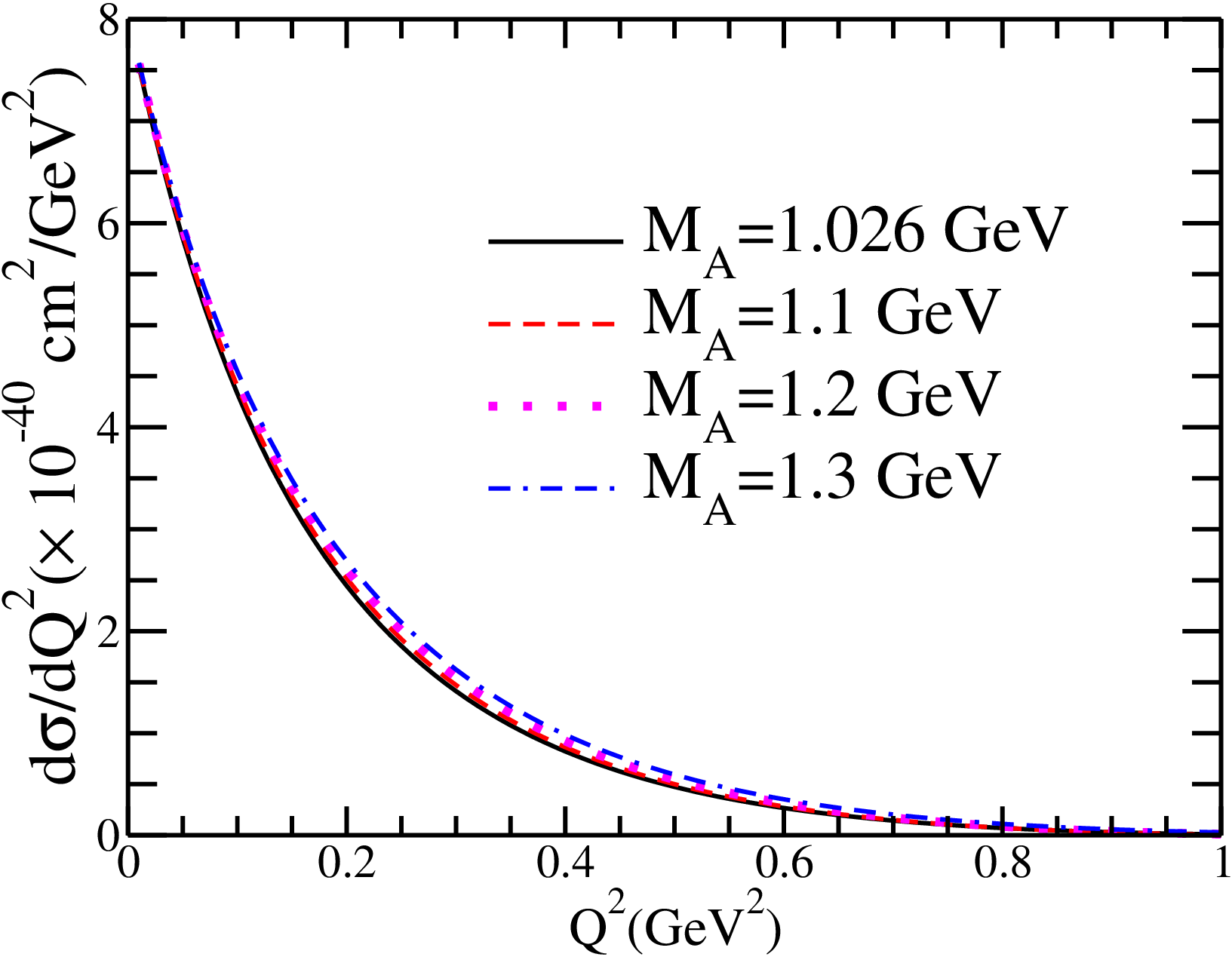}
   \includegraphics[height=5cm,width=5cm]{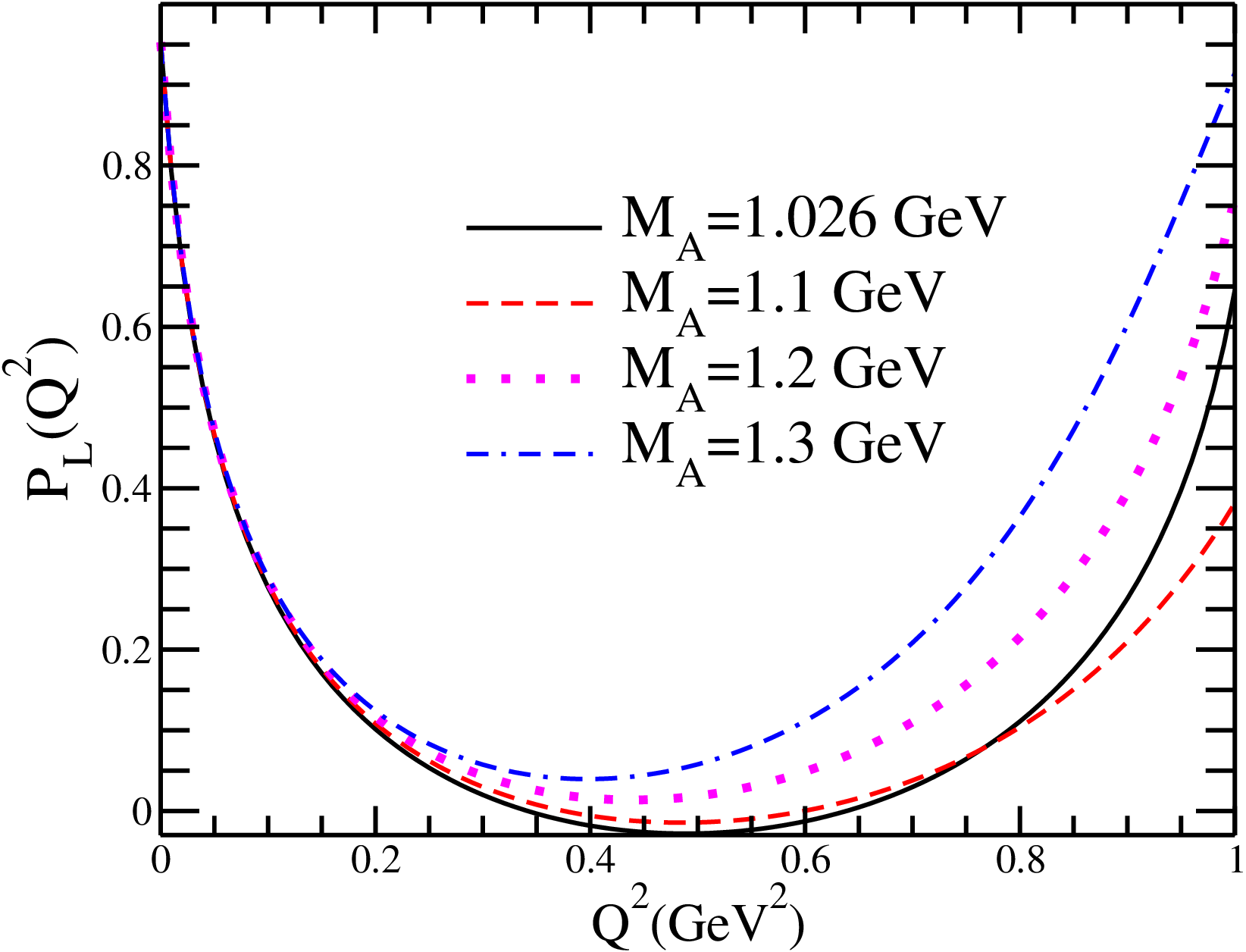}
   \includegraphics[height=5cm,width=5cm]{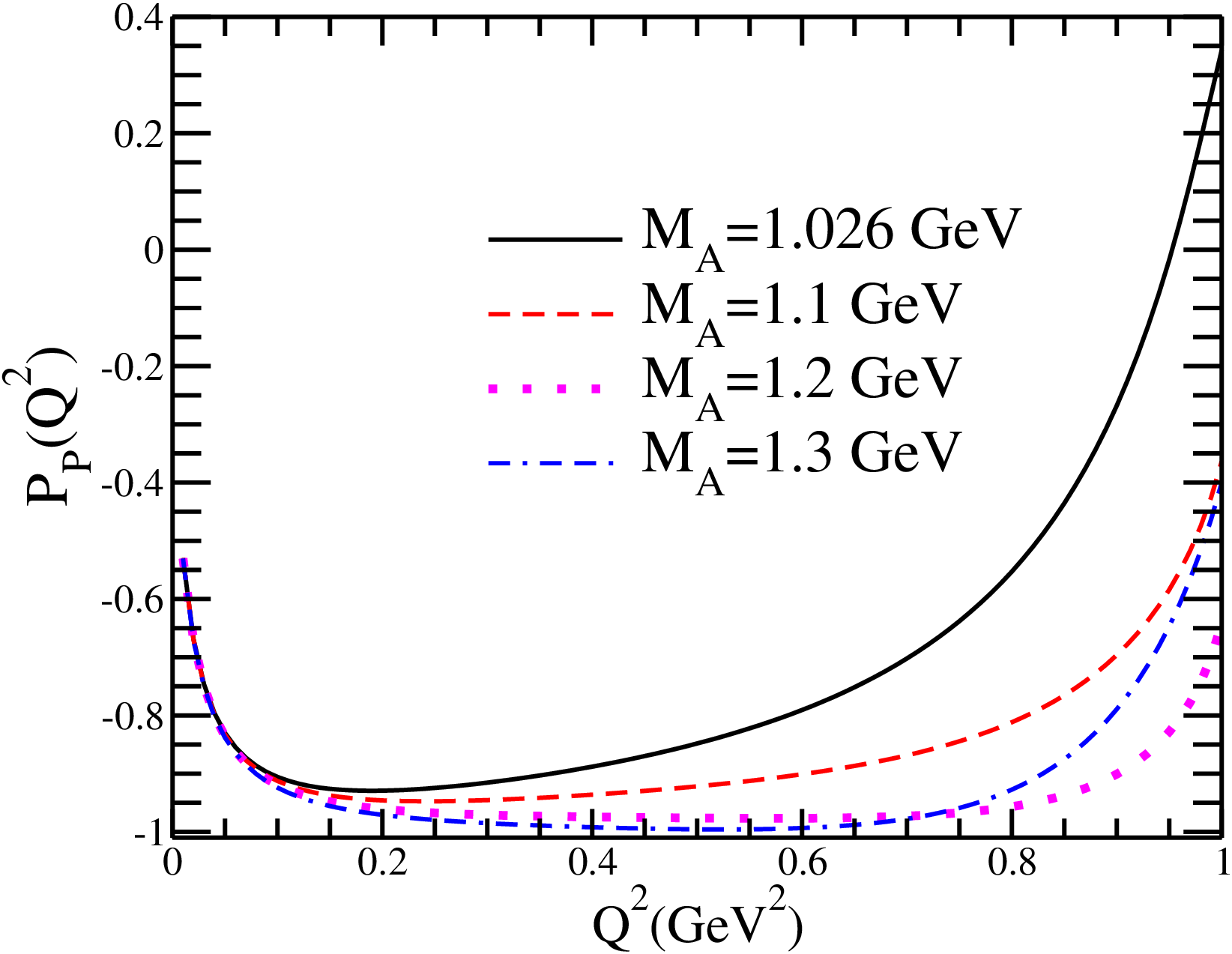}
    \caption{ $\frac{d\sigma}{dQ^2}$, $P_L(Q^2)$ and $P_P(Q^2)$ vs $Q^2$ for the process $\bar \nu_\mu p \to \mu^+ \Lambda$ at $E_{\bar \nu_{_\mu}}$ = 1 GeV 
   for different values of $M_A$ used in $g_1^{p\Lambda}(Q^2)$ 
    viz. 1.026 GeV(solid line), 1.1 GeV(dashed),
   1.2 GeV(dotted line) and 1.3 GeV(double dashed-dotted line)  with  $m_\mu=0 $.}
\label{Ma_dstbn}
  \end{figure}
\subsection{Polarization of hyperons}
Using the covariant density matrix formalism, polarization 4-vector($\xi^\tau$) of the final hyperon produced
in Eq.~(\ref{hyp-rec}) is written as:
\begin{equation}\label{polar}
\xi^{\tau}=\frac{\mathrm{Tr}[\gamma^{\tau}\gamma_{5}~\rho_{f}(p^\prime)]}
{\mathrm{Tr}[\rho_{f}(p^\prime)]},
\end{equation}
where the final spin density matrix $\rho_f(p^\prime)$ is given by 
\begin{equation}\label{polar1}
 \rho_{f}(p^\prime)= {\cal L}^{\alpha \beta}  \Lambda(p')J_{\alpha} \Lambda(p)\tilde{J}_{\beta} 
\Lambda(p').
\end{equation} 

Using the following relations~\cite{Bilenky:2013fra}

$$\Lambda(p')\gamma^{\tau}\gamma_{5}\Lambda(p')=2M_Y\left(g^{\tau\sigma}-
\frac{p'^{\tau}p'^{\sigma}}{M_Y^{2}}\right)\Lambda(p')\gamma_{\sigma}
\gamma_{5};~~
 \Lambda(p^\prime)\Lambda(p^\prime) = 2M_Y \Lambda(p^\prime),$$
 $\xi^\tau$ defined in Eq.~\ref{polar} may be rewritten as:
 \vspace{-5mm}
\begin{equation}\label{polar4}
\xi^{\tau}=\left( g^{\tau\sigma}-\frac{p'^{\tau}p'^{\sigma}}{M_Y^2}\right)
\frac{  {\cal L}^{\alpha \beta}  \mathrm{Tr}
\left[\gamma_{\sigma}\gamma_{5}\Lambda(p')J_{\alpha} \Lambda(p)\tilde{J}_{\beta} \right]}
{ {\cal L}^{\alpha \beta} \mathrm{Tr}\left[\Lambda(p')J_{\alpha} \Lambda(p)\tilde{J}_{\beta} \right]}.
\end{equation}
Note that in Eq.~\ref{polar4}, $\xi^\tau$ is manifestly orthogonal to $p^{\prime \tau}$ i.e. $p^\prime \cdot \xi=0$. Moreover, the denominator
is directly related to the 
differential cross section 
\begin{equation}\label{dsig}
 \frac{d\sigma}{dQ^2}=\frac{G_F^2 \sin^2\theta_c}{8 \pi M E_{\bar \nu_{_\mu}}^2} {\cal N}(Q^2,E_{\bar \nu_{_\mu}}),
\end{equation} 
where the expression of ${\cal N}$ is given in the appendix of Ref.~\cite{Akbar:2016awk}.

With ${\cal J}^{\alpha \beta}$ and ${\cal L}_{\alpha \beta}$ obtained using hadronic and leptonic currents defined in  Eq.\ref{lephad}, $\xi^\tau$ is evaluated. In the lab frame where the initial nucleon 
is at rest, the polarization vector $\vec \xi$ is calculated to be 
\begin{equation}\label{3pol}
\frac{d\sigma}{dQ^2} \vec \xi =\frac{G_F^2 \sin^2 \theta_c }{8 \pi\; M M_Y E^2_{\bar \nu_\mu}}\left[(\vec k + \vec k^{\prime})M_Y {\cal A}(Q^2,E_{\bar \nu_{_\mu}}) + (\vec k -
\vec k^{\prime}) {\cal B}(Q^2,E_{\bar \nu_{_\mu}}) \right], 
\end{equation}
where the expressions of ${\cal A}(Q^2,E_{\bar \nu_{_\mu}})$ and ${\cal B}(Q^2,E_{\bar \nu_{_\mu}})$ are given in the
appendix of Ref.~\cite{Akbar:2016awk}.
 \begin{figure}
   \includegraphics[height=5cm,width=5cm]{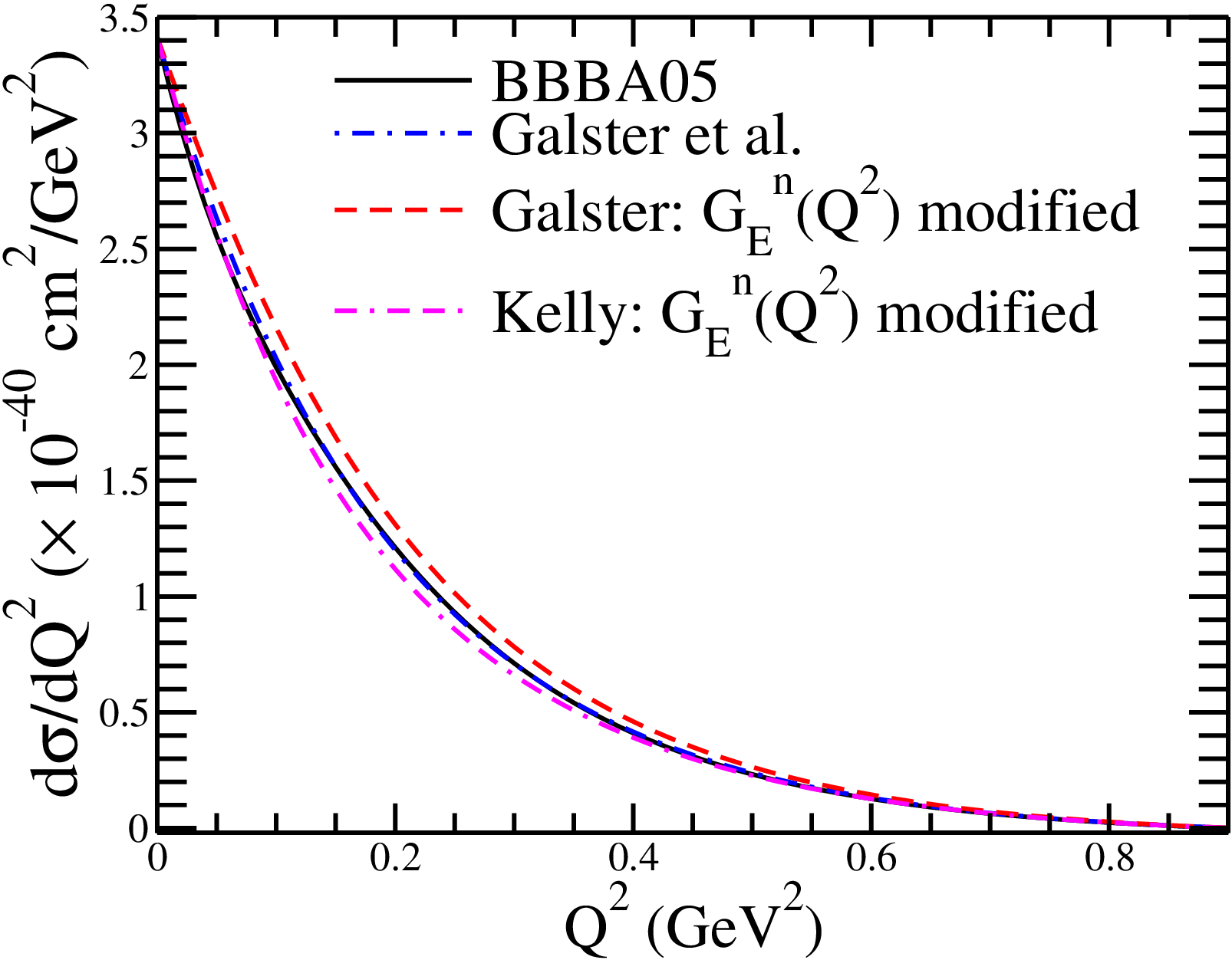}
   \includegraphics[height=5cm,width=5cm]{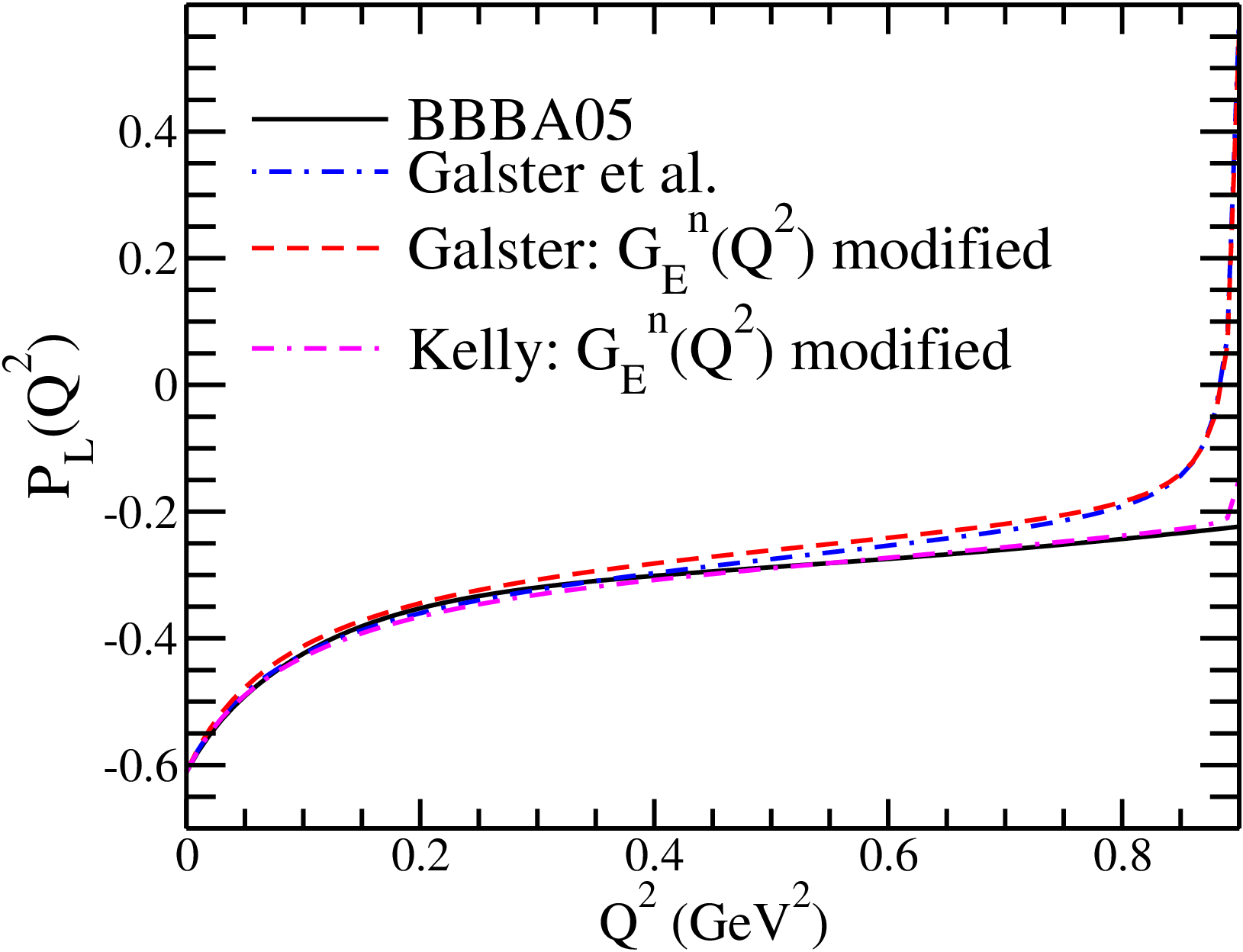}
   \includegraphics[height=5cm,width=5cm]{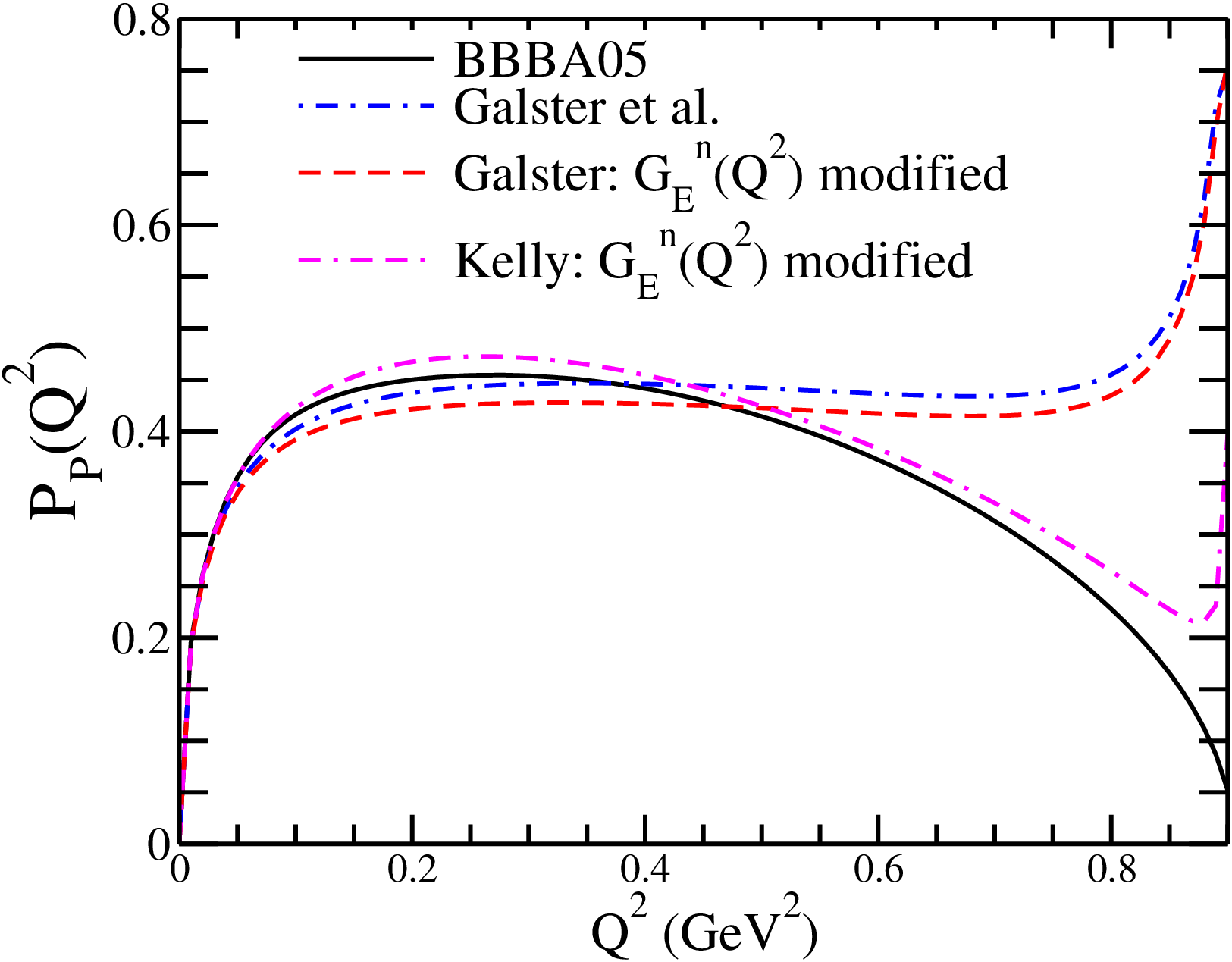}
   \caption{$\frac{d\sigma}{dQ^2}$, $ P_L(Q^2)$ and $P_P(Q^2)$ vs $Q^2$ at $E_{\bar \nu_{_\mu}}$= 1 GeV for $\bar\nu_\mu n \to \mu^+ \Sigma^-$ process.
The results are presented with the 
  nucleon form factors using BBBA05~[16](solid line), Galster et al.~[17]
  (dashed-dotted line), modified form of $G_E^n(Q^2)$ 
  in Galster parameterization~[18](dashed line) and modified form of $G_E^n(Q^2)$ in
  Kelly parameterization~[19](double dashed-dotted line).}\label{sig_gen}
\end{figure}

\begin{figure}
   \includegraphics[height=5cm,width=5cm]{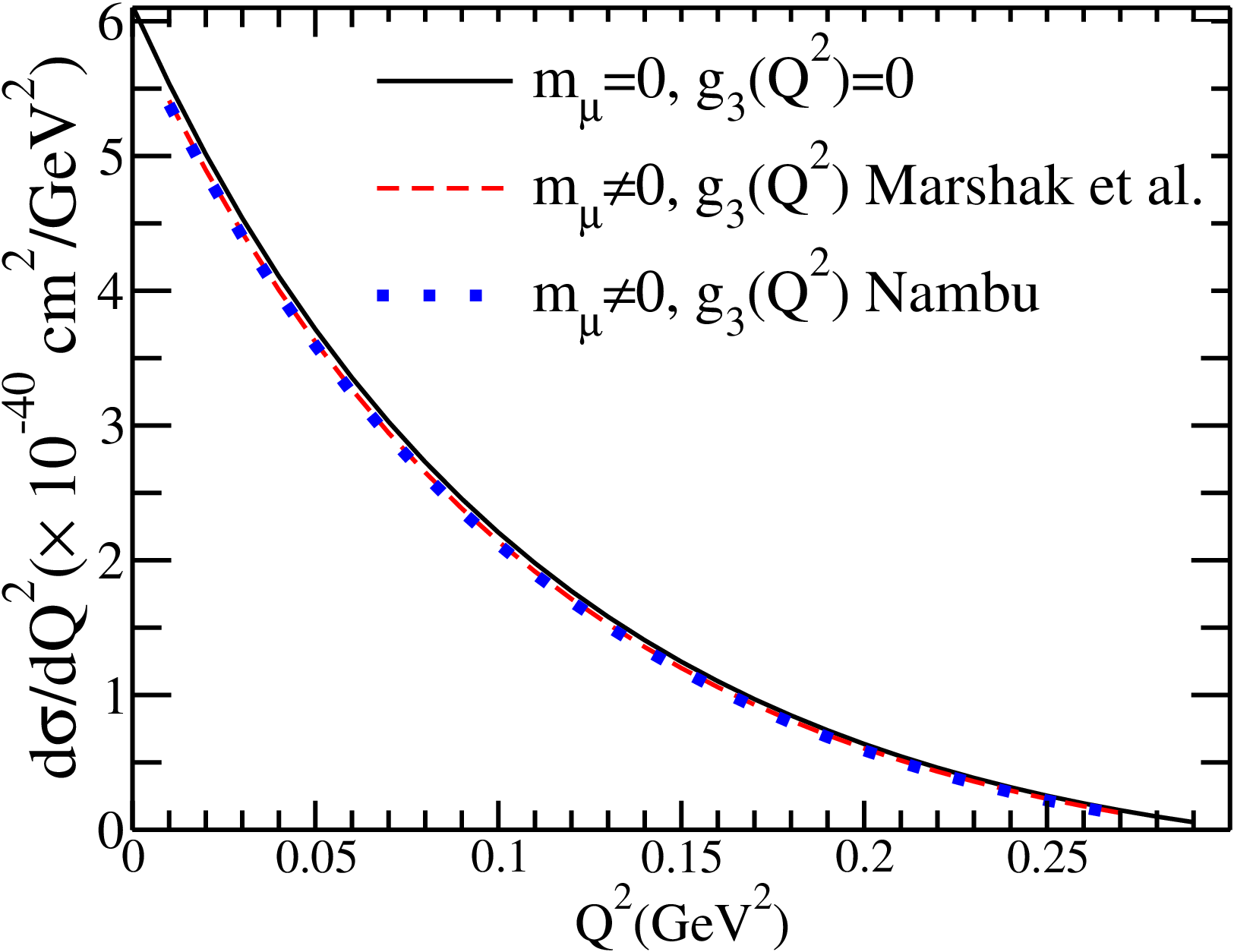}
   \includegraphics[height=5cm,width=5cm]{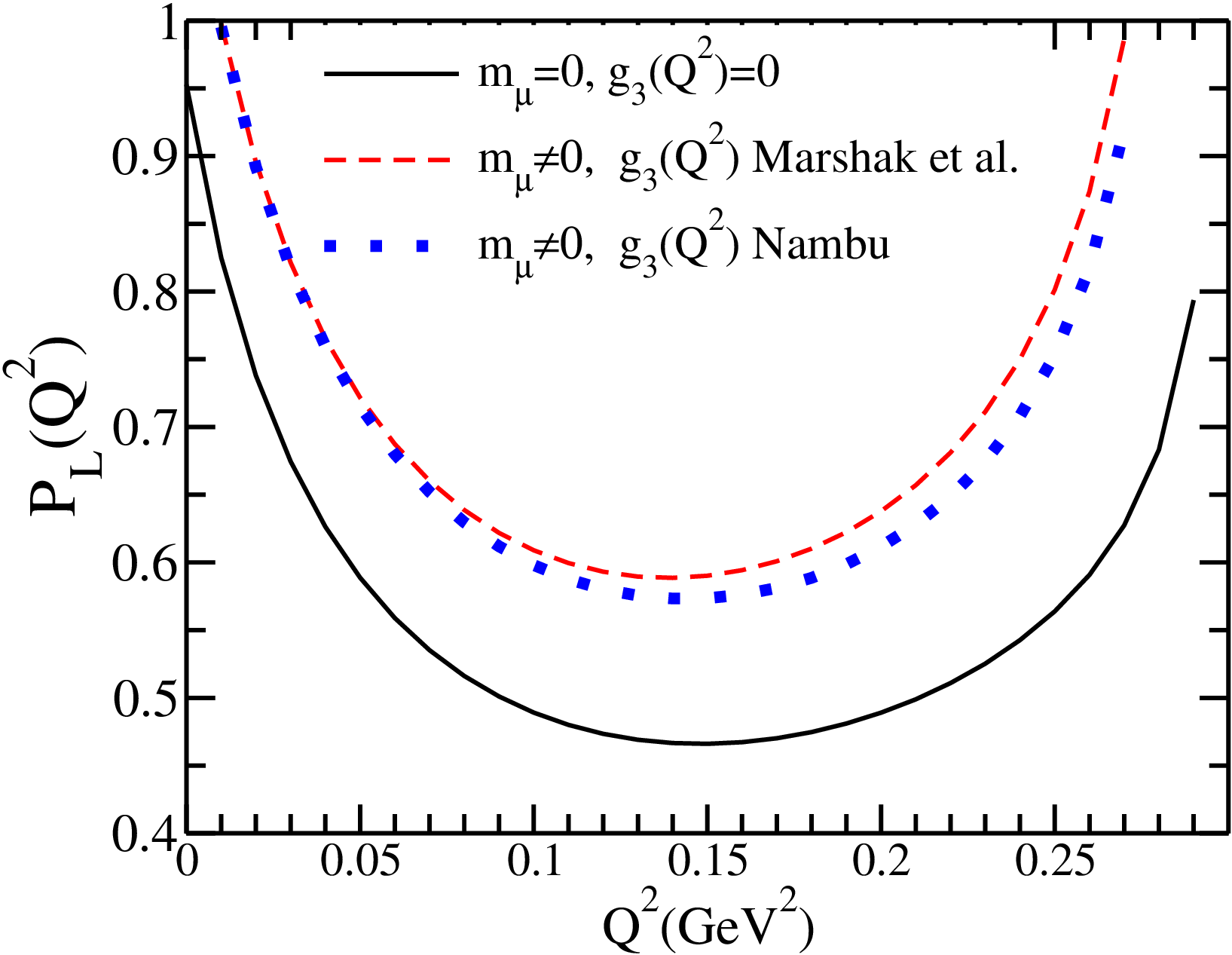}
   \includegraphics[height=5cm,width=5cm]{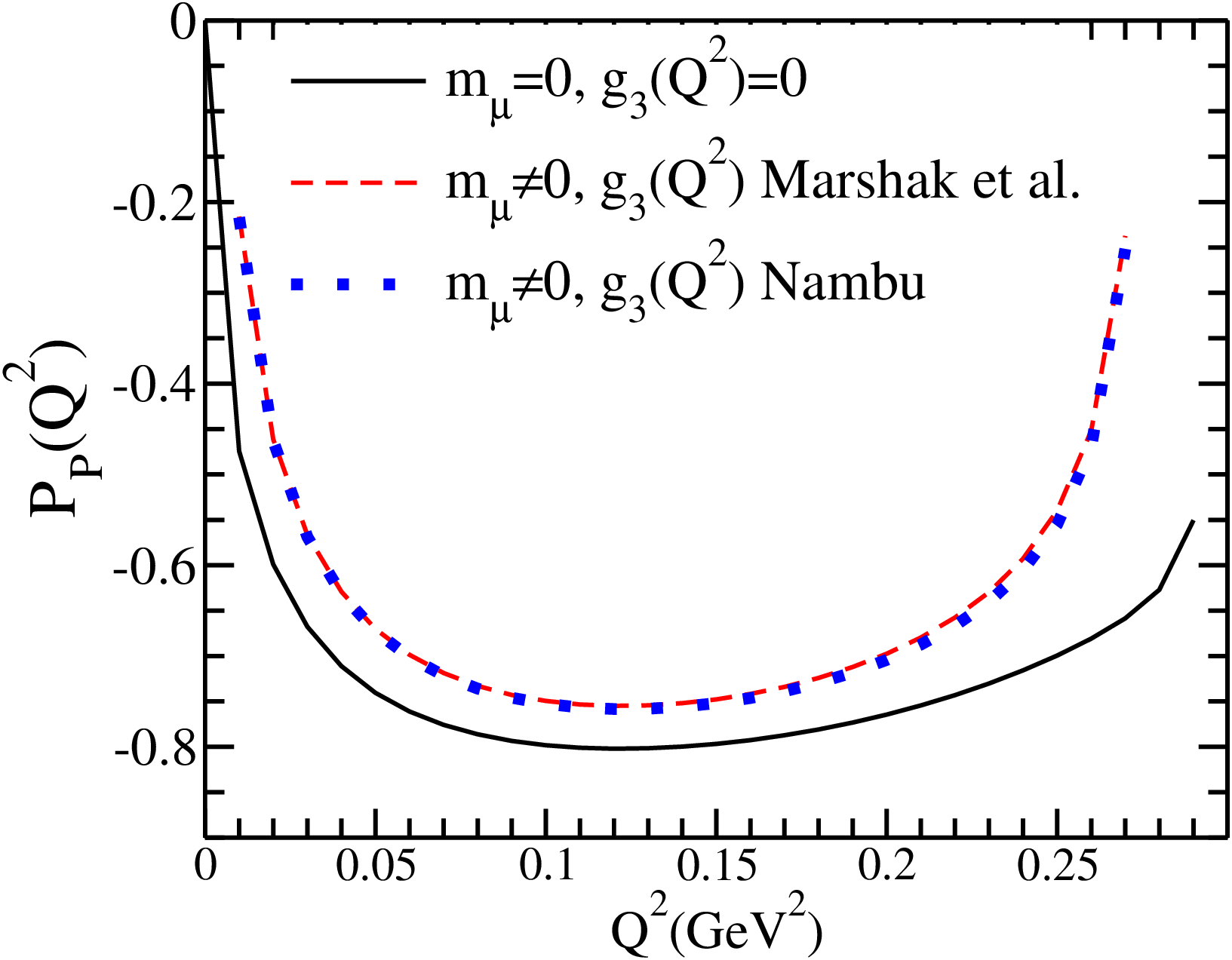}
   \caption{ $\frac{d\sigma}{dQ^2}$, $P_L(Q^2)$ and $P_P(Q^2)$ vs $Q^2$ ($M_A=$ 1.026 GeV)
   for the process $\bar \nu_\mu p \to \mu^+ \Lambda$ at $E_{\bar \nu_\mu}=~0.5$ GeV
   using $f_1^{p\Lambda}(Q^2),~f_2^{p\Lambda}(Q^2),~g_1^{p\Lambda}(Q^2)$ 
   from Table-I of Ref.[3],  BBBA05~[16] parameterization for the nucleon form factors,
    with $m_\mu = 0$ and $g_3^{p\Lambda} = 0$(solid line), 
$m_\mu \ne 0$ and $g_3^{p\Lambda} \ne 0$ from Marshak et al.~[14](dashed line) and 
$m_\mu \ne 0$ and $g_3^{p\Lambda} \ne 0$ from Nambu~[15](dotted line).}\label{fp_vartn}
\end{figure}
From Eq.~\ref{3pol}, it follows that the polarization vector lies in scattering plane defined by $\vec k$ and $\vec k^\prime$, and there is no component of
polarization in the direction orthogonal to the scattering 
plane. This is a consequence of T--invariance which makes the transverse polarization in a direction perpendicular to the reaction plane to
vanish. We now expand the polarization vector $\vec \xi$ along two orthogonal directions, $\vec e_L$ and $\vec e_P$ in the reaction plane
corresponding to parallel and perpendicular directions to the momentum of hyperons given by
\begin{equation}\label{vectors}
\vec{e}_{L}=\frac{\vec{p'}}{|\vec{p'}|}=\frac{\vec{q}}{|\vec{q}|},\qquad
\vec{e}_{P}=\vec{e}_{L}\times \vec{e_T},\qquad  \vec{e_T}=\frac{\vec{k}\times\vec{k'}}{|\vec{k}\times\vec{k'}|},
\end{equation}
and write
 \begin{equation}\label{polarLab}
\vec{\xi}=\xi_{P}\vec{e}_{P}+\xi_{L}\vec{e}_{L},\qquad \xi_L(Q^2)=\vec \xi \cdot \vec e_L,\qquad \xi_P(Q^2)=\vec \xi \cdot \vec e_P.
\end{equation}
From Eq.~\ref{polarLab}, the longitudinal and perpendicular components of polarization vector $P_L(Q^2)$ and $P_P(Q^2)$ defined in the rest frame of recoil nucleon are given by ~\cite{Bilenky:2013fra}:
\begin{equation}\label{PL1}
 P_L(Q^2)=\frac{M_Y}{E_{p^\prime}} \xi_L(Q^2), \qquad P_P(Q^2)=\xi_P(Q^2),
\end{equation}
where $\frac{M_Y}{E_{p^\prime}}$ is the Lorentz boost factor along $\vec p^\prime$.
With the help of Eqs.~\ref{3pol}, \ref{vectors}, \ref{polarLab} and \ref{PL1}, the longitudinal component $P_L(Q^2)$ is calculated to be
\begin{equation}\label{sl}
\frac{d\sigma}{dQ^2} P_L(Q^2)= \frac{G_F^2 \sin^2\theta_c}{8\pi|\vec q| E_{p^{\prime}} M\;E^2_{\bar \nu_{_\mu}}} \left[\left(E_{\bar \nu_{_\mu}}^2 - E_\mu^2 + m_\mu^2 \right)M_Y
  {\cal A}(Q^2,E_{\bar \nu_{_\mu}}) + |\vec q|^2 {\cal B}(Q^2,E_{\bar \nu_{_\mu}})\right],
\end{equation}
where in the lab frame $E_{p^{\prime}} = \sqrt{|{\vec q}^2|+M_Y^2}$. 
Similarly, the perpendicular component $P_P(Q^2)$  of the polarization 3-vector is given as
\begin{equation}\label{st}
\frac{d\sigma}{dQ^2} P_P(Q^2) = -\frac{G_F^2 \sin^2\theta_c}{4\pi} \frac{|\vec k^{\prime }|}{|\vec q|}\frac{{\cal A}(Q^2,E_{\bar \nu_{_\mu}}) \sin\theta}{M E_{\bar \nu_{_\mu}}},
\end{equation}
where $\theta$ is the scattering angle in the lab frame.
\subsection{Results}
 In Fig.~\ref{fg:xsec_ar}, we present the results of $\sigma(E_{\bar \nu_\mu})$ vs $E_{\bar \nu_\mu}$ for $\Lambda$ and $\Sigma^-$ productions
 in $^{40}Ar$ and $^{56}Fe$. We find that nuclear medium effects due to Pauli blocking are very small. 
 However, the final state interactions due to $\Sigma-N$ and $\Lambda - N$ interactions in various channels 
 tend to increase the $\Lambda$ production and decrease the $\Sigma^-$ production. 
 The quantitative increase(decrease) in $\Lambda(\Sigma)$ yield due to FSI increases with the increase in nucleon 
number. The $\Sigma^-$ and $\Sigma^0$
production are separately affected and the relation 
$ \sigma \left(  \bar \nu_\mu + p \rightarrow \mu^+ + \Sigma^0   \right) = 
\frac12  \sigma \left(  \bar \nu_\mu + n \rightarrow \mu^+ + \Sigma^-   \right) $
is modified in the nucleus due to the presence of other nucleons. 

 In  Fig.\ref{difel.fig}, we have presented the results for the $Q^2$ distribution averaged over the
MiniBooNE~\cite{AguilarArevalo:2013hm} and Super-Kamiokande~\cite{Honda:2006qj} spectra. We must point out that
$\pi^-$ production from hyperon excitations is scaled by 2.5. The results are presented for the
differential cross sections calculated with nuclear medium and  pion absorption effects for the pions obtained 
from the decay of $\Delta$ excitation.  We observe that, in the peak region of $Q^2$ distribution, the contribution of $\pi^-$ from 
 the hyperon excitations is almost 40$\%$ to the
 contribution of $\pi^-$ from the $\Delta$ excitation.
 
 In Fig.~\ref{Ma_dstbn},
 we present the results of $\frac{d\sigma}{dQ^2}$, $P_L(Q^2)$ and $P_P(Q^2)$ for the reaction
$\bar\nu_\mu p \to \mu^+ \Lambda$
 at $E_{\bar \nu_{_\mu}}=1$ GeV. We observe that 
 while there is very little sensitivity of $\frac{d\sigma}{dQ^2}$ to the variation of $M_A$, the components of 
polarization $P_L(Q^2)$ and $P_P(Q^2)$ are 
sensitive to the value of $M_A$ specially in the region $Q^2>0.4$ GeV$^2$.
It should, therefore, be possible to independently determine the value of $M_A$ from the
polarization measurements even though the presently
available data on the total cross section for the single hyperon production are
consistent with $M_A=1.026$ GeV. At higher values of $Q^2$, the sensitivity of $P_L(Q^2)$ and $P_P(Q^2)$ to $M_A$ increases,
 but quantitatively, the cross section $\frac{d\sigma}{dQ^2}$ decreases making the
number of events quite small and the measurement of polarization observables becomes difficult.

 We show in Fig.~\ref{sig_gen}($E_{\bar \nu_{_\mu}}=1$ GeV), the dependence of $\frac{d\sigma}{dQ^2}$,
$P_L(Q^2)$ and $P_P(Q^2)$ on the different parameterization of $G_E^n(Q^2)$ for $\Sigma^-$. It 
 should be possible to determine, in principle, the charge form factor of neutron from the observation of $P_L(Q^2)$ and $P_P(Q^2)$ using this process.
 
 We have made an attempt to explore the possibility of determining the pseudoscalar form factor $g_3^{NY}(Q^2)$ in 
 $|\Delta S|=1$ sector by including two models for $g_3^{NY}(Q^2)$ based on PCAC  
 and the corresponding Goldberger--Treiman relation in the strangeness sector using the parameterizations
 given
 by Marshak et al.~\cite{Marshak} and Nambu~\cite{Nambu:1960xd},
 respectively. In Fig.~\ref{fp_vartn},
 we show the effect of $g_3^{NY}(Q^2)$ on $\frac{d\sigma}{dQ^2}$,  $P_L(Q^2)$ and $P_P(Q^2)$ calculated for the processes
  $\bar\nu_\mu p \to \mu^+\Lambda$ at $E_{\bar \nu_{_\mu}}$=0.5 GeV.
\vspace{-5mm}
 \section{Conclusion}
 \vspace{-5mm}
\begin{itemize}
\item The reduction due to nuclear medium and FSI effects in the case of pions obtained from $\Delta$ excitation is large enough to compensate
for Cabibbo suppression of pions produced through hyperon excitations up to $E_{\bar\nu_\mu}<0.5 $GeV for $\pi^-$ production and 0.65 GeV for $\pi^0$ production.
\vspace{-3mm}
\item Pion production from hyperon excitation when folded with the spectra of T2K or atmospheric antineutrino fluxes found to be significant.
\vspace{-3mm}
\item These results may have important applications in the analysis of MicroBooNE, MINER$\nu$A, ICARUS and SBND experiments.
\vspace{-3mm}
\item Polarization observables $P_L(Q^2)$ and $P_P(Q^2)$ are quite sensitive to the value of $M_A$ for $Q^2>0.4$ GeV$^2$ and $M_A$ may be measured independent of $\sigma$.
\vspace{-3mm}
\item $Q^2$ dependence of cross sections and polarization components are found to be sensitive to the $G_E^n(Q^2)$ especially for $\Sigma^- $.
\vspace{-3mm}
\item $P_L(Q^2)$ and $P_P(Q^2)$ are sensitive to the pseudoscalar form factor at low $E_{\bar\nu_\mu}$.
\end{itemize}
%

%

\end{document}